%Paper: 9202003
%From: kamion@guinness.ias.edu (Marc Kamionkowski)
%Date: Sun, 2 Feb 92 18:57:21 EST
%Date (revised): Thu, 6 Feb 92 18:25:22 EST

%this is axion.tex (final version)
\input phyzzx
\nonstopmode
\sequentialequations
\twelvepoint
\nopubblock
\tolerance=5000
\overfullrule=0pt

\REF\QCD{For a review of QCD see F.~Wilczek, {\it Ann. Rev.
Nucl. Part. Sci.} {\bf 32}, 177 (1982).}

\REF\peccei{R.~Peccei and H.~Quinn, {\it Phys. Rev. Lett.} {\bf
38}, 1440, (1977); {\it Phys. Rev. D} {\bf 16}, 1791 (1977).}

\REF\axion{F.~Wilczek, {\it Phys. Rev. Lett.} {\bf 40}, 279 (1978);
S.~Weinberg, {\it Phys. Rev. Lett.} {\bf 40}, 223 (1978).  For a
review of axions see, for example, J.~E.~Kim, {\it Phys. Rep.}
{\bf 150}, 1 (1987).}

\REF\us{M. Kamionkowski and J.~March-Russell, Institute for
Advanced Study Report No. IASSNS-HEP-92/6, Princeton University
Report No.
PUPT-92-1304; see also  R.~Holman, S.~Hsu, E.~W.~Kolb, R.~Watkins,
and L.~M.~Widrow, UC Santa Barbara, Institute for Theoretical
Physics Preprint.}

\REF\gs{S. Giddings and A. Strominger, {\it Nucl. Phys.} {\bf B307},
854 (1988).}

\REF\coleman{S. Coleman, {\it Nucl. Phys.} {\bf B310}, 643 (1988).}

\REF\gilbert{G. Gilbert, {\it Nucl. Phys.} {\bf B328}, 159 (1989). }

\REF\nohair{See, for example, B. Carter in {\it General
Relativity.  An Einstein Centenary Survey,} edited by
S.~W.~Hawking and W.~Israel (Cambridge University Press,
Cambridge, 1979).}

\REF\hawking{S. W. Hawking, {\it Comm. Math. Phys.} {\bf 43},
199 (1975).}

\REF\astro{See, for example, M.~S.~Turner, {\it Phys. Rep.} {\bf
197}, 67 (1991); G.~Raffelt, {\it Phys. Rep.} {\bf 198}, 1 (1990);
E.~W.~Kolb and M.~S.~Turner, {\it The Early Universe}
 (Addison-Wesley, Redwood City, 1990).}

\REF\witten{See for instance: M.~Green, J.~Schwarz, and E.~Witten,
{\it Superstring Theory, Vol. 2} (Cambridge University Press,
Cambridge, 1987).}

\REF\shenker{S. Shenker, {\it The Strength of Nonperturbative Effects
in String Theory}, Rutgers Report No. RU-90-47 (1990).}

\REF\seckel{S.~M.~Barr and D.~Seckel, University of Delaware,
Bartol Research Institute Preprint.  We thank David Seckel for
notifying us of this work.}

\REF\progress{M.~Kamionkowski and J.~March-Russell, work in
progress.}

\REF\discrete{L.~M.~Krauss and F. Wilczek, {\it Phys. Rev. Lett.}
{\bf 62}, 1221 (1989).
M. Alford, S. Coleman, and J. March-Russell, {\it Nucl.
 Phys.} {\bf B351}, 735 (1991); M. Alford, J. March-Russell, and
 F. Wilczek, {\it Nucl. Phys.} {\bf B337}, 695 (1990); J.
 Preskill, L.~M.~Krauss, {\it Nucl. Phys.} {\bf B341}, 50 (1990).}

\REF\old{H. Georgi, L. Hall and M. Wise, {\it Nucl. Phys.}
{\bf B192}, 409 (1981). G.~Lazarides, C.~Panagiotakopoulos
and Q.~Shafi, {\it Phys. Rev. Lett.} {\bf 56}, 432 (1986).
We thank S.~Hsu, E.~Kolb and D.~Seckel for informing us of these
papers.}

\REF\itp{R.~Holman, S.~Hsu, E.~W.~Kolb, R.~Watkins, and L.~M.~Widrow,
UC Santa Barbara, Institute for Theoretical Physics Preprint.}

% Some useful abbreviations:
\def\fun#1#2{\lower3.6pt\vbox{\baselineskip0pt\lineskip.9pt
  \ialign{$\mathsurround=0pt#1\hfil##\hfil$\crcr#2\crcr\sim\crcr}}}
\def\lap{\mathrel{\mathpalette\fun <}}
\def\gap{\mathrel{\mathpalette\fun >}}
\def\order{{\cal O}}

\def\Th{\bar{\th}}
\def\Mp{M_{\rm Pl}}

\let\de=\delta

\let\th=\theta

\let\om=\omega

\let\<=\langle
\let\>=\rangle

\def\comment#1{ \hbox{Comment suppressed here.} }
\def\tr{\hbox{tr}\,}
\def\fpq{{f_{PQ}}}

\line{\hfill IASSNS-HEP-92/9}
\line{\hfill PUPT-92-1309}
\line{\hfill January 1992}
\titlepage
\title{Planck-Scale Physics and the Peccei-Quinn Mechanism}
\medskip
\author{Marc Kamionkowski\foot{Research supported
by an SSC Fellowship from the Texas National Research
Laboratory Commission. e-mail: kamion@guinness.ias.edu}}
\smallskip
\centerline{{\it School of Natural Sciences}}
\centerline{{\it Institute for Advanced Study}}
\centerline{{\it Olden Lane}}
\centerline{{\it Princeton, N.J. 08540}}
\medskip
\author{John March-Russell\foot{Research supported by NSF
grant
NSF-PHY-90-21984. e-mail: jmr@puhep1.princeton.edu,
jmr@iassns.bitnet}}
\smallskip
\centerline{{\it Joseph Henry Laboratories}}
\centerline{{\it Princeton University}}
\centerline{{\it Princeton, N.J. 08544}}
\medskip

\abstract{Global-symmetry
violating higher-dimension operators, expected to be induced by
Planck-scale physics, in general drastically alter the properties
of the axion field associated with the Peccei-Quinn solution to
the strong-$CP$ problem, and render this solution unnatural.
The particle physics and cosmology associated with other global
symmetries can also be significantly changed.}

\endpage

After almost
twenty years of experimental verification, there is little
room to doubt that quantum chromodynamics (QCD) is the true
theory of the strong interactions [\QCD].  Perhaps the only
outstanding
flaw in the the theory arises from non-perturbative effects
which, unless suppressed, lead to a neutron electric-dipole
moment orders of magnitude larger than that observed.  This is
the infamous strong-$CP$ problem.  Essentially, the problem is that
the QCD Lagrangian contains a term
$$
{\Th} {g^2\over32\pi^2} G^{a\mu\nu}\widetilde{G}_{a\mu\nu},
\eqn\term
$$
where $G^{a\mu\nu}$ is the gluon field and $\Th$ is an
undetermined parameter.  This term leads to an electric-dipole
moment of order $d_n\simeq 5\times10^{-16}\Th$ e cm.  The
current experimental limit is $d_n\lap10^{-25}$ e cm which
constrains ${\Th}$ to be less than $10^{-10}$.
Here we have performed an anomalous chiral rotation to move
the phase of the determinant of the fermion mass-matrix into the
theta-term, resulting in a net theta-angle ${\Th}$.

To date, the most elegant and intriguing solution to the
strong-$CP$ problem has been that proposed by Peccei and Quinn
[\peccei] where ${\Th}$ becomes
a dynamical field with a potential
minimized at ${\Th}=0$.  Their solution involves introducing a new
global chiral symmetry $U(1)_{PQ}$ spontaneously broken at a
scale $\fpq$ which leads to a Nambu-Goldstone boson, the
axion [\axion].  Due to the anomalous nature of the $U(1)_{PQ}$ symmetry,
QCD-instanton (and  other,  more general, non-perturbative)
effects  result  in  the axion acquiring a periodic potential
$$
V_{\rm QCD}(\Th)=(m_a^i)^2 \fpq^2(1-\cos\Th),
\eqn\potential
$$
minimized at $\Th=0$ (where, for simplicity, we consider the case where
no axion domain walls occur).
Here
$$
m_a^i\simeq 0.4 {f_\pi m_\pi \over \fpq}.
\eqn\amass
$$
is the mass of the axion induced by QCD non-perturbative effects.

In this Letter we make the simple observation that the existence
of higher-dimension symmetry-violating operators expected to be
induced at the Planck scale by quantum-gravity effects spoils the
Peccei-Quinn solution to the strong-$CP$ problem.  Generally, the explicit
Planck-scale symmetry-violating effects will favor a minimum of
the potential at a value $\Th\neq0$.  In order for the
Peccei-Quinn mechanism to work the symmetry-breaking effects from
Planck-scale physics must be small compared to those
induced by QCD effects which drive $\Th$ to zero.  What
we find is that in order to solve the strong-$CP$ problem, either;
(1) the couplings of symmetry-breaking operators from the Planck
scale must be exponentially small, or (2) the Planck-scale
potential is coincidentally minimized at $\Th=0$.  Therefore,
the Peccei-Quinn mechanism, invoked to solve a ``fine-tuning''
problem, is itself subject to a similar disease.

To reprise the arguments of
Ref.~\us, it is widely believed that Planck-scale
physics results in the violation of all global symmetries.
Wormholes provide one specific mechanism for this violation
[\gs,\coleman,\gilbert].  Black holes provide yet another.  It
is well known
that as a consequence of the black-hole no-hair theorems [\nohair]
the global charge of a black hole is not defined; therefore, if
in a scattering process a virtual or non-virtual black hole is
formed from an initial state of definite global charge, the
black hole decays (Hawking evaporates)
[\hawking] into final states of differing
global charge. At energies small compared to the Planck mass these
symmetry-violating effects may be described by higher-dimension
operators in an effective theory of the light modes. On
dimensional grounds, the higher dimension operators are expected
to be suppressed by the appropriate power of the Planck mass
resulting in symmetry-breaking operators like those introduced
in Eq. (5) below.

The calculation is simple.  The potential for the $U(1)_{PQ}$
field $\phi$ is
$$
V_0(\phi)=\lambda(|\phi|^2-\fpq^2/2)^2.
\eqn\mexican
$$
First we add to this a general explicit symmetry-breaking term,
that might well be induced by quantum-gravity effects, of
dimension $2m+n$ and $U(1)_{PQ}$ charge of $n$:
$$
V_g(\phi) = {g\over \Mp^{2m+n-4}} |\phi|^{2m} \phi^n + h.c. +
c \, ,
\eqn\DelV
$$
where $g=|g|\exp(i\de)$ is a complex coupling which might well be of
order unity, and $c$ is a
constant chosen so that the minimum of $V$ is zero.
Note that we are {\it not} necessarily
assuming that quantum-gravity effects
explicitly violate $CP$. In the case where they do
not violate $CP$ (as we expect), the phase $\delta$ is
just proportional to $\arg\det M_f$, and arises from the
chiral $U(1)_{PQ}$ rotation that we perform to move
the phase of the fermion mass matrix into the $\th$-term.

After spontaneous-symmetry breaking, the potential for the axion
degree of freedom $a$ due to Planck-scale effects becomes
$$
V_g=(m_a^g)^2\fpq^2[1-\cos(na+\delta)],
\eqn\Plpot
$$
where we define the square of the
quantum-gravitationally induced axion mass to be
$$
(m_a^g)^2=|g|{\Mp^2}\left({\fpq\over\sqrt{2}\Mp}\right)^{2m+n-2}.
\eqn\mag
$$
When we add this to the potential due to QCD instanton effects,
Eq.~\potential, the complete potential for $a$ becomes
$$
V(a)=\fpq^2\left\{(m_a^g)^2[1-\cos(na+\delta)] + (m_a^i)^2
[1-\cos (a)] \right\},
\eqn\complete
$$
where we have shifted the axion field $a$ so
as to eliminate the $\Th  G\widetilde{G}$ term
in the action, and implicitly defined a new $\de$.
(The full calculation would involve the running
of the coupling constant $g$ down from the Planck scale to
the scale of the axion mass. This leads to additional factors
expected to be of order $\ln(\Mp/m_a)\lap 50$ in $g$ which, as
we will see, does not alter our main conclusions. However
such considerations can be important in more general contexts.)

In any case,
in order for the Peccei-Quinn mechanism to solve the strong-$CP$
problem, the minimum of the potential should be located at
$a\lap10^{-10}$.  After some algebra, this
condition (taking $n=1$ for simplicity) may be written as
$$
{|\sin\delta|\over (1+r^2+2r\cos\delta)^{1/2}} \lap 10^{-10},
\eqn\condition
$$
where $r\equiv (m_a^i)^2/(m_a^g)^2$.  Therefore, if $\sin\delta$ is
or order unity, we must have $r\gap10^{10}$.

If we demand that the coupling constant $|g|$ be
$\order(10^{-2})$
and assume a symmetry-breaking operator of dimension 5 we find
that in order to consistently solve the strong-$CP$ problem we
must have
$$
\fpq\lap10\, {\rm GeV},
\eqn\ruleout
$$
which corresponds to axion masses $m_a\gap 100$ keV.  Axions with
such masses coming purely from the anomaly and strong-interaction
physics have been ruled out by {\it laboratory}
experiments [\astro].  Since in this regime $m_a^g\ll
m_a^i$, the standard phenomenology of the axion will
remain essentially unaltered by the Planck-scale physics (in
other words the coupling of the axion to other fields is
determined by $\fpq $ which in this regime is almost unaltered from
its usual value for these masses).  Therefore these
laboratory results remain valid, and an axion that is both able
to naturally solve the strong-$CP$ problem, and is affected
by Planck-scale physics in the assumed way,
is disallowed by observation. This is our most significant
conclusion.

Of  course,  we could take the couplings of the
symmetry-violating operators to be
exponentially small.  For instance,
astrophysical arguments have ruled out most possible values of
the axion mass [\astro] (assuming that the mass is related to
the matter couplings in the
standard way).  The only open window is now around
$m_a^i\sim10^{-5}$ eV, which corresponds to $\fpq\sim10^{12}$ GeV.
If the Peccei-Quinn mechanism is to work with such a
symmetry-breaking scale, the coupling of a dimension-5
symmetry-breaking operator induced by Planck-scale physics must
be
$$
|g|\lap 10^{-55}.
\eqn\gcond
$$
Note that for this value of $g$ it is self-consistent to
use unamended astrophysical arguments since the couplings are
essentially unchanged. But all we have achieved is to replace
one mystery, the smallness of the observed value of $\Th$,
with another, the smallness of the coupling constants $g$.
(Or alternatively, the mystery of why
their phases $\de$ are inexplicably
related to the phase of the determinant of the fermion mass
matrix, an apparently low-energy phenomenon).

Now, of course, axion-like fields have been considered in the
context of string theory.\foot{We thank Tom Banks and Ed Witten
for discussions on these points.}  For instance, the
``model-independent'' axion arises from the two-form field
$B_{\mu\nu}$, and the characteristic axion couplings to
$\tr G\tilde{G}$ (and $\tr R\tilde{R}$, where $R$ is the
Ricci curvature) result from the Yang-Mills and Lorentz
Chern-Simons three forms $\om_Y$ and $\om_L$, that appear
in field strength $H=dB-\om_Y+\om_L$. However, not only
do QCD non-perturbative effects explicitly break the Peccei-Quinn
symmetry, but so do string-theoretic non-perturbative
effects [\witten].
The important point is that, unlike Yang-Mills theories where
instanton effects are generally of order $\exp(-8\pi^2/e^2)$,
string instanton effects are of order $\exp(-2\pi/e)$
[\shenker]. A reasonable (and expected) order of magnitude
value of the string coupling is $e\sim e_{GUT}\sim 0.5$,
leading to a value of
the coupling of the higher-dimension operator $g\sim\exp(-15)$.
It is significant that our constraint on the coupling,
$g\lap 10^{-55}\sim \exp(-130)$, is stronger than this value.

We should point out that if some mechanism {\it does}
select $\Th=0$ as the minimum of the potential, then
the Peccei-Quinn mechanism will still work.  However, if
$m_a^g\gg m_a^i$ the mass of the axion will be $m_a\simeq
m_a^g$, and for a given value of $\fpq$ the
phenomenology will be altered significantly [\seckel,\progress].
In short, since the QCD-instanton induced mass is a decreasing
function of $\fpq$ and the quantum-gravitationally induced mass
is an increasing function of $\fpq$, there will in general be a
minimum mass for the axion.  For example, assuming a dimension-5
symmetry-breaking operator and $|g|\sim10^{-2}$ (at the axion
scale), the minimum mass
is roughly $m_a\sim 1{\rm keV}$ and occurs for
$\fpq\sim10^4$ GeV.
Astrophysics and cosmology have been used to severely restrict
the allowed values of $\fpq$ [\astro]; however, if the axion has
a gravitationally induced mass, the constraints on $\fpq$ need
to be re-examined and several are most likely invalidated
[\seckel,\progress].

Another point is that one
might imagine a situation in which the coefficients of the
leading symmetry-violating operators are very small (or zero),
but those of some much higher-dimension (\eg, dimension 40,000)
symmetry-violating
operators are large.  For instance if we take $\fpq\sim10^{12}$
GeV (for an astrophysically allowed solution to the strong-$CP$
problem) and assume that  the first operator with a large
coupling has $|g|$ of order $10^{-2}$, we find that the leading
operator consistent with the Peccei-Quinn solution has dimension
$2m+n\simeq12$. However, this is an erroneous conclusion, unless
there is some reason (such as protected continuous, or discrete
[\discrete], gauge symmetries) that {\it forbids} all the
lower-dimension operators. The reason for this is
simple:\foot{We thank Sidney Coleman for this argument.}  If the
effective theory contains a symmetry-violating operator of the form
$|\phi|^{2m}\phi^n$ with a coupling of order unity, then we can form
symmetry-violating operators of lower dimension $n$ by
contracting legs. The inverse powers of $\Mp$ from the operators
are cancelled by positive powers of $\Mp$ from the divergent
loop integrations that must be cut off at the scale of new
physics (in our case, the Planck mass), leaving suppression only
by the dimensionally enforced power of $\Mp$ and a
coupling that by assumption is not small.  More generally, if we
have two operators, both with couplings or order unity, that
violate global charge
by amounts $n$ and $n'$ (and have arbitrary values of $m$),
we can then form an operator of dimension
$|n-n'|$ that also violates global charge (unless, of course,
such an operator is ``accidentally'' forbidden by some other reason,
such as gauge invariance). Thus, the only
consistent  ways in which Planck-scale symmetry violations can
be suppressed are: (a) by having the lower-dimension operators
absolutely forbidden, or (b) by having {\it all} of the couplings
of the symmetry-violating operators exponentially small.

In summary, we have shown that, unless suppressed, higher-dimension
symmetry-violating operators induced by quantum-gravity effects
generally drive $\Th$ to a value other than 0, invalidating
the Peccei-Quinn solution to the strong-$CP$ problem.  On the
other hand, if for some reason the Planck-scale physics picks
out $\Th=0$, the phenomenology of the standard axion is
significantly altered and the astrophysical constraints on
$\fpq$ may not be as restrictive as currently believed.

More generally, what we have essentially argued is that for each
(continuous or discrete) global symmetry spontaneously broken at
a scale $f$, Planck-scale physics induces a characteristic
explicit-symmetry breaking scale resulting in a mass
$m$ for the pseudo-Nambu-Goldstone boson given roughly by
$$
m^2 \sim gn^2 f^2 \left({f \over \Mp}\right)^{2m+n-4}.
\eqn\scale
$$
Since many ideas in particle physics and cosmology rely on
exact or (nearly exact) global symmetries, it is clear that
the existence of higher-dimensional global-symmetry-violating
operators can have significant
consequences, for example, on the texture [\us] and late-time
phase-transition models
for large-scale structure formation, the evolution of
global cosmic strings and monopoles, some baryogenesis
scenarios,  various candidate explanations for the dark matter,
several recently-proposed inflationary models,
and particle-physics models involving majorons, familons,
schizons, or spontaneously broken discrete global symmetries
(which have been thought to be severly constrained
cosmologically by the evolution of domain walls)
[\progress].  Viewed from this perspective, global
symmetries are a significant, and stringent, test of the physics
of the Planck scale, with many phenomenological ramifications.

After the completion of this work we were informed that
previous authors have commented upon the effects of
Planck-scale physics on the properties of the axion [\old].
We also understand that similar conclusions have been reached by
R.~Holman, S.~Hsu, E.~W.~Kolb, R.~Watkins, and L.~M.~Widrow
[\itp], and by D.~Seckel and S.~M.~Barr [\seckel].  We
gratefully thank  Tom Banks, Robert Brandenburger,
Sidney Coleman, Jacques Distler, Jerry Michael, John Preskill,
David Seckel, Erick Weinberg, Frank Wilczek and Ed Witten for
helpful discussions.  MK gratefully acknowledges the hospitality
of the Institute for Advanced Study.

\par

\refout

\end